\documentstyle[aps,multicol]{revtex}

\newcommand{\bef}{\begin{figure}}
\newcommand{\enf}{\end{figure}}
\newcommand{\bec}{\begin{center}}
\newcommand{\enc}{\end{center}}
\newcommand{\PSbox}[3]{\mbox{\rule{0in}{#3}\includegraphics{#1}\hspace{#2}}}

\newcommand{\Vds}{V_{\mbox{\scriptsize ds}}}
\newcommand{\Vg}{V_{\mbox{\scriptsize g}}}
\newcommand{\GL}{\Gamma_{\mbox{\scriptsize L}}}
\newcommand{\GR}{\Gamma_{\mbox{\scriptsize R}}}
\newcommand{\dbg}{\delta_0}
\newcommand{\dres}{\delta_{\mbox{\scriptsize res}}}

\newcommand{\Gbg}{G_{\mbox{\scriptsize inc}}}
\newcommand{\kB}{k_{\mbox{\scriptsize B}}}

\newcommand{\Pcorr}{\Phi_{\mbox{\scriptsize corr}}}
\newcommand{\gd}{g_{\mbox{\scriptsize 2D}}}

\begin{document}
\bibliographystyle{prsty}
\title{Fano Resonances in Electronic Transport through a Single Electron
Transistor}
\author{J. G\protect{\"{o}}res\cite{joernaddress}, D. Goldhaber-Gordon\cite{davidaddress},
S. Heemeyer, and M.A. Kastner\cite{byline}}
\address{
Department of Physics, Massachusetts Institute of Technology\\
Cambridge, MA 02139
}
\author{Hadas Shtrikman, D. Mahalu, and U. Meirav}
\address{
Braun Center for Submicron Research\\
Weizmann Institute of Science\\
Rehovot, Israel 76100
}

\maketitle
\vspace{.1in}
\begin{abstract}

We have observed asymmetric Fano resonances in the conductance of a
single electron transistor resulting from interference between a
resonant and a nonresonant path through the system. The resonant
component shows all the features typical of quantum dots, but the
origin of the non-resonant path is unclear. A unique feature of this
experimental system, compared to others that show Fano line shapes, is
that changing the voltages on various gates allows one to alter the
interference between the two paths.  
\end{abstract}


\begin{multicols}{2} \narrowtext \section{Introduction}

When a droplet of electrons is confined to a small region of space
and coupled only weakly to its environment, the number of electrons 
and their energy become quantized. The
analogy between such a system and an atom has proved to be quite
close. In particular, these artificial atoms exhibit
properties typical of natural atoms, including a charging energy for
adding an extra electron and an
excitation energy for promoting confined electrons to higher-lying
energy levels~\cite{kastner,ashoori}. Remarkably, the analogy goes
even further and includes cases where an artificial atom interacts
with its environment. A system known as a single electron transistor
(SET), in which an artificial atom is coupled to conducting
leads, can be accurately described by the Anderson model~\cite{ng,glazman,meir}.
The same model has been used extensively to study the interaction
of localized electrons with delocalized ones within a metal containing
magnetic impurities. One of its subtle predictions is the Kondo effect,
which involves many-body correlations between an electronic spin on an
impurity atom and those in the surrounding metal. This effect has 
recently
been observed in an SET when the artificial atom develops a net spin
because of an odd electron occupancy~\cite{david,davidprl,cronenwett,jschmidt,simmel}.
In this paper we report that by changing the parameters in a single
electron transistor we observe another phenomenon typical of natural
atoms: Fano resonances.
While several aspects of the Fano resonances in our SETs
are easily understood, others are very surprising.

Asymmetric Fano lineshapes are ubiquitous in resonant
scattering~\cite{fano,feshbach,shibatani}, and
are observed whenever resonant and nonresonant scattering paths
interfere. The more familiar symmetric Breit-Wigner or Lorentzian
lineshape~\cite{breit} is a limiting case that
occurs when there is no interference, for example when there is no
non-resonant scattering channel. Fano resonances have been observed in a
wide variety of experiments including atomic photoionization
\cite{fano2}, electron and neutron scattering~\cite{adair,simpson},
Raman scattering~\cite{cardona}, and photoabsorption in quantum well
structures~\cite{capasso,schmidt}.
\bef[htb]
\centering
\PSbox{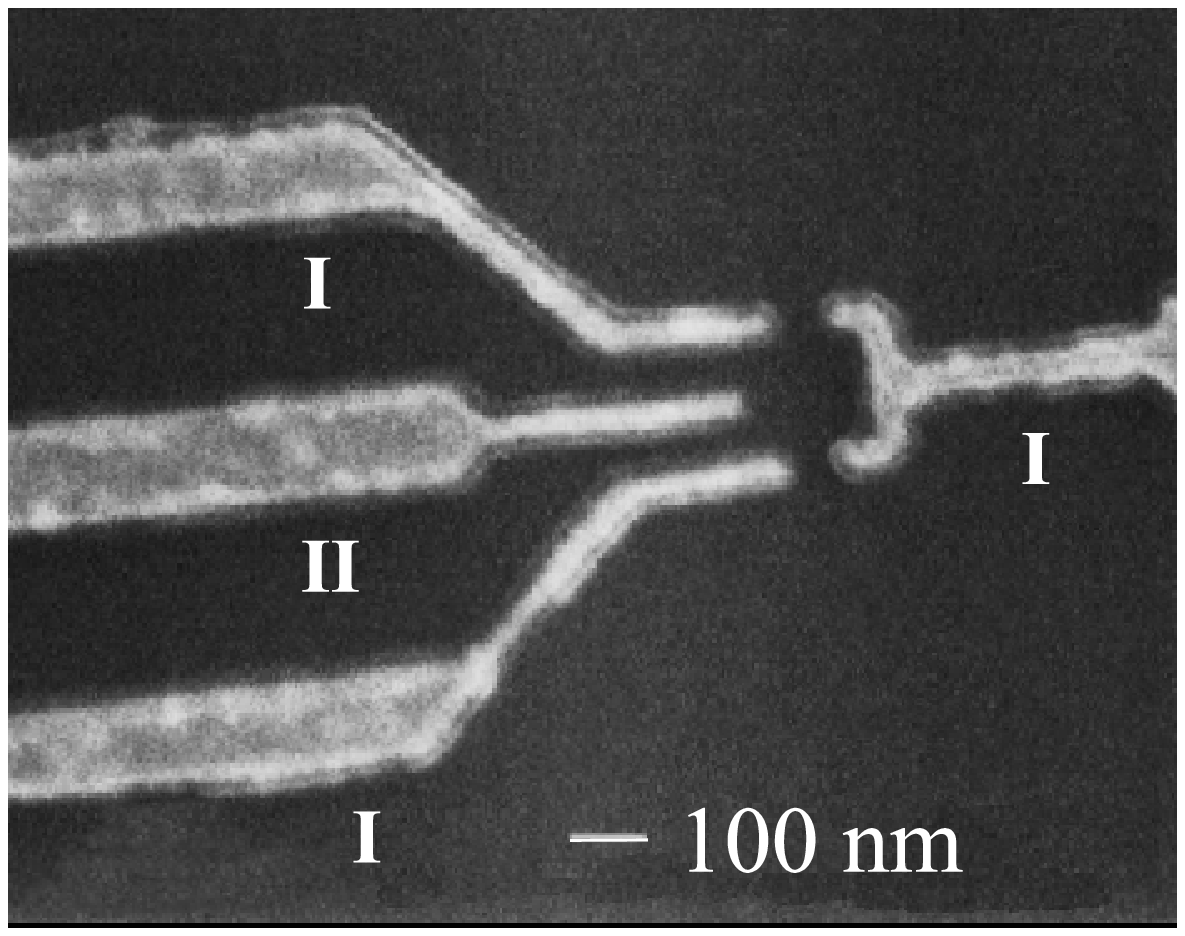 hscale=45 vscale=45}{5.5cm}{4.5cm}
\caption{Electron micrograph of
an SET showing the split gates (I) that define the tunnel barriers
and the additional gate electrode (II) that adjusts the potential
energy on the droplet.}
\label{figure1}
\enf
The widely successful application of the Landauer-B\"{u}ttiker formalism
\cite{landauer,buettiker} shows that electron transport through a
mesoscopic system is closely analogous to scattering in the systems
described above. Therefore, one might expect to also observe Fano
resonances in the conductance of 
nanostructures~\cite{noeckel,porod,tekman,price}.
Indeed, Fano lineshapes have been reported in a recent experiment by Madhavan 
$\it{et\ al.}$ measuring
tunneling from an STM tip through an impurity atom into a metal surface
\cite{madhavan}.
However, this characteristic feature of interference between
resonant and
non-resonant scattering has not been reported for more conventional
nanostructures.

In this paper we report the observation of Fano resonances in 
conductance
through a single electron transistor (SET). This structure has the
advantage
over the tunneling experiment of Madhavan $\it{et\ al.}$ \cite{madhavan} and over conventional
scattering experiments that we can tune the key parameters and thus study the interference leading to Fano
resonances in greater detail.
This paper is organized as follows: In section II we describe the SET and the measurements we have made on it; we also summarize there the standard theory for Fano line shapes. Our results are presented in section III. In section IV we discuss the results, draw conclusions and point out issues which require further research.

\section{Experimental details and theoretical background}

An SET consists of a small droplet of electrons coupled to two 
conducting
leads by tunnel barriers. Gate electrodes (shown in Fig.~\ref{figure1})
are used to control the electrostatic potential of the droplet and, in our
structures, the heights of the tunnel barriers. The SETs used in these
experiments are the same ones used for studies by Goldhaber-Gordon
$\it{et\ al.}$ of the Kondo effect~\cite{david,davidprl}. The electron
droplet is
created in a two-dimensional electron gas (2DEG) that forms at
the interface of a GaAs/AlGaAs heterostructure with a mobility of 
$100,000\ \mbox{cm}^2/\mbox{Vs} $ and a density of 
$8.1\times 10^{11}\ \mbox{cm}^{-2}$. Applying a negative
voltage to two pairs of split gate
electrodes depletes the 2DEG underneath them and
forms two tunnel barriers. The barriers separate the droplet of
electrons from the 2DEG regions on either side, which act as the
leads. The heights of the two barriers can be adjusted independently by
changing the voltages on the respective constriction
electrodes (I), and the potential energy of the electrons on the droplet can
be shifted relative to the Fermi energies
in the leads using an additional plunger gate electrode (II) near the droplet.

Our SET's are made with a 2DEG that is closer to the surface
($\approx 20$ nm) than in most other experiments, allowing the electron
droplet to be confined to smaller dimensions. This also makes the tunnel
barriers more abrupt than in previous structures. We estimate that 
our droplet is about $100$ nm in diameter, smaller than the lithographic
dimensions because of depletion, and contains about $50$ electrons.

For conductance measurements we apply a small alternating drain-source
voltage (typically $5\mu$V) between the leads and measure the pre-amplified current 
with a lock-in amplifier. The conductance is then recorded
as a function of the plunger gate voltage $\Vg$. For differential conductance
measurements we add a finite offset drain-source voltage $\Vds$
and measure the response $dI/d\Vds$ to the small alternating
drain-source voltage as a function of both $\Vg$ and $\Vds$.

In an SET the Coulomb
interaction among electrons opens up an energy gap $U$ in the tunneling
spectrum, given classically by $e^2/2C$, where $C$ is the total 
capacitance
between the the droplet and its environment, primarily the nearby
conducting leads and gates. Thus, an
energy $U$ is required to overcome Coulomb repulsion and add an electron to
the
droplet. This energy cost causes the number of electrons on
the droplet to be quantized and electron transport through the droplet 
to
be suppressed. However, lowering the chemical potential of the 
droplet by adjusting the voltage on the plunger gate makes it possible 
to add electrons one at a time. At a charge degeneracy
point, where the states with $N$ and $N+1$ 
electrons on the droplet have equal energy, transport of electrons from
one lead through the
droplet to the other lead is allowed.
This effect is known as Coulomb blockade~\cite{fultondolan}
because transport is
suppressed everywhere except close to the degeneracy points.
Because of the small size of the electron droplet in our SETs, the 
energy
spacing between the discrete levels
$\Delta\epsilon$, that is, the energy to excite the droplet at fixed $N$,
is only a few times smaller than
the charging energy $U$.

Depending on the transmission of the left and right tunnel barriers,
characterized by 
tunneling rates $\GL/h, \GR/h$, respectively, we observe
different transport regimes in our SETs at very low temperature.
If the thermal energy $\kB T$ is much smaller than
the coupling energy $\Gamma \equiv \GL+\GR$,
quantum fluctuations dominate so that the resonances have width $\Gamma$. When $\Gamma\ll\Delta\epsilon$ the coupling is weak and one observes narrow quasi-periodic peaks (Fig.~\ref{figure2}(c)).
As the coupling is increased a new regime emerges, in which transport
away from the resonances is enhanced by the Kondo effect~\cite{david} when 
there is an odd number of electrons on the droplet (Fig.~\ref{figure2}(b)). 
Surprisingly, if we increase the coupling beyond the 
Kondo regime we observe asymmetric Fano resonances on top of a slowly 
varying background (Fig.~\ref{figure2}(a)). Before discussing these data in detail, we 
present the analytic form predicted for such line shapes.

Scattering theory, specifically the S-matrix approach, predicts Fano
lineshapes as the general form for resonances in transport through
quasi-1D systems \cite{noeckel}. The S-matrix, which relates the outgoing and incoming
scattering state amplitudes, is a unitary matrix at real
energies $\epsilon$ because of probability conservation.
Therefore, for real energies the eigenvalues of S can be written
in the form $\lambda(\epsilon) = e^{2i\delta(\epsilon)}$ where
$\delta(\epsilon)$ is the scattering phase. 

Resonant scattering occurs when one of the eigenvalues of the
S-matrix develops a pole at the complex energy $\epsilon_0 - i\Gamma/2$. Near
this pole the resonant eigenvalue is given by $\lambda(\epsilon) =
e^{2i\dbg}e^{2i\delta_{\mbox{\tiny res}}(\epsilon)}$~\cite{noeckel}.
Here $\dres(\epsilon)$ is the resonant contribution to the
phase shift and is given by
$\tan\dres(\epsilon) = -\Gamma/2(\epsilon-\epsilon_0)$. It 
varies
from zero to $\pi$ as the energy is moved through the
resonance from below. The background or nonresonant contribution to the
scattering phase shift $\dbg$ is, in this approximation,
a constant independent of the energy.

The total cross section is directly related to the scattering phase shifts through
\begin{equation}
\sigma(\epsilon) \propto \sin^2(\dres(\epsilon) + \dbg).
\end{equation}
\bef[htb]
\PSbox{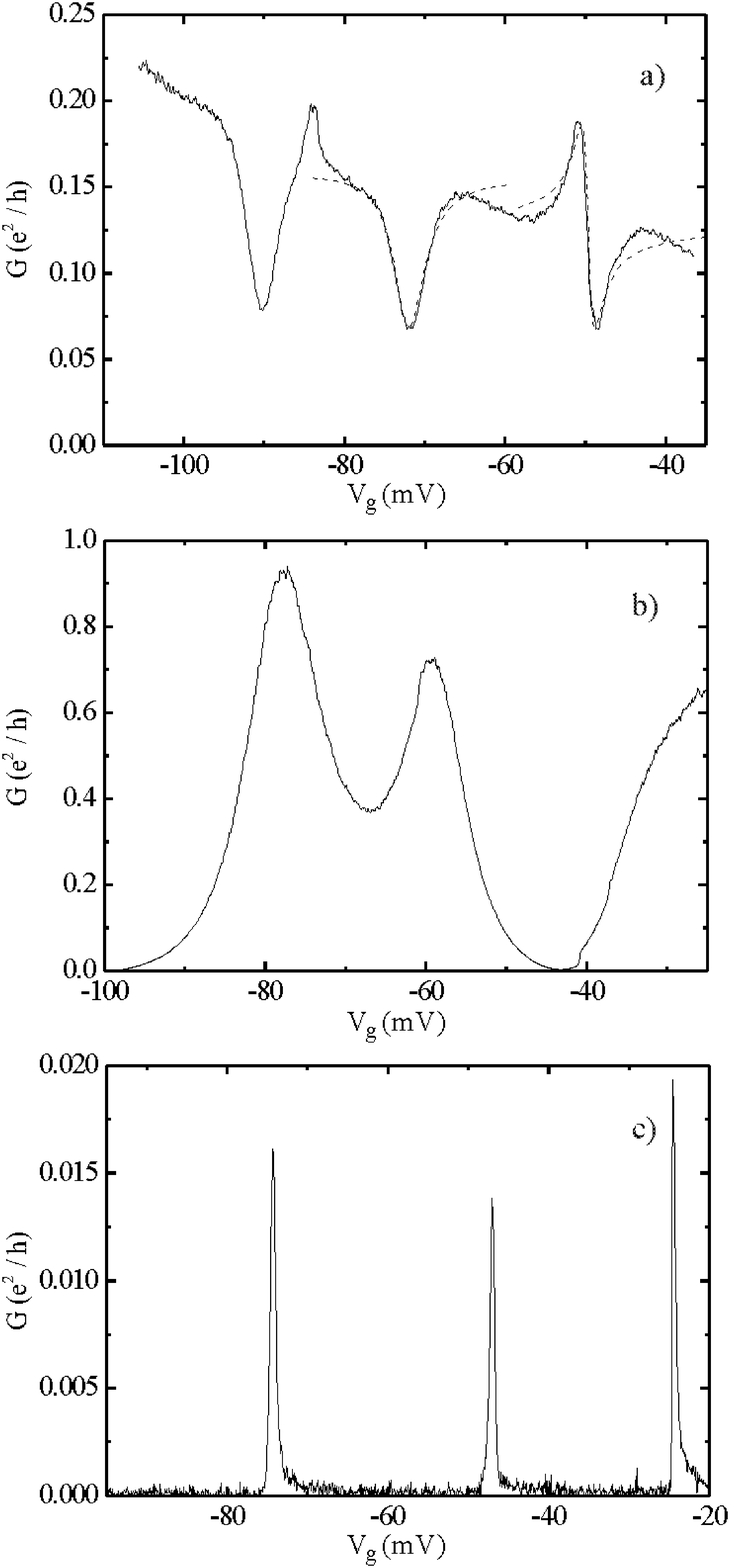 hscale=30 vscale=30}{14cm}{17.3cm}
\centering
\caption{Comparison of conductance measurements in the (a) Fano regime, 
(b) Kondo regime [6] and (c) weak coupling regime [6]. Fits to the Fano formula (\ref {qformula2}) are shown for the center and right resonances in (a). The respective asymmetry parameters are $q = -.03$ and $q = -.99$. During the sweep through the left resonance a sudden shift in effective gate voltage occured, presumably caused by charge motion in the vicinity of the droplet, leading to the unusual shape of this resonance.}
\label{figure2} 
\enf
\noindent Varying the value of the background phase shift produces the entire
family of Fano lineshapes.

It is important to emphasize that two interfering
channels are necessary for Fano resonances to arise. One is resonant, 
for
which the phase changes by $\pi$ in an energy interval $\approx\Gamma$
around the resonance energy $\epsilon_0$. The other is nonresonant
and has a constant background phase shift
$\dbg$. If there is no
nonresonant channel or the background phase shift between the
channels is zero, symmetric Breit-Wigner resonances are 
observed. In all other cases Fano resonances result.

In his original work Fano treated the case of inelastic scattering in 
the context of autoionization and derived the so-called Fano formula for
the scattering cross section~\cite{fano}
\begin{equation}
\sigma(\epsilon) \propto
\frac{(\tilde{\epsilon}+q)^2}{\tilde{\epsilon}^2+1},
\label{qformula}
\end{equation}  
where $\tilde{\epsilon} \equiv (\epsilon - \epsilon_0)/(\Gamma/2)$ is 
the
dimensionless detuning from resonance and $q$ is
called the asymmetry parameter. The asymmetry parameter is related to the background phase
shift of the S-matrix treatment by $q = \cot\,\dbg$. 
The magnitude of $q$ is proportional to the ratio of transmission amplitudes for the resonant and non-resonant channels \cite{fano}. The limit $q \rightarrow \infty$, in which resonant transmission dominantes, leads to symmetric Breit-Wigner resonances. In the opposite limit $q \rightarrow 0$ non-resonant transmission dominates, resulting in symmetric dips. 

According to the Landauer-B\"{u}ttiker formalism, conductance through 
any mesoscopic system can be expressed in terms of an
S-matrix. Hence, Fano resonances should also be observed in conductance if a
resonant and a nonresonant transmission path coexist~\cite{noeckel}. Analogous to the scattering case Eq.~(\ref{qformula}) the conductance is then given by 
\begin{equation}
G = \Gbg + G_0\frac{(\tilde{\epsilon}+q)^2}{\tilde{\epsilon}^2+1},
\label{qformula2}
\end{equation}
where $\Gbg$ denotes an incoherent contribution to the conductance, which is often observed\cite{qcomplex}. Note that the Breit-Wigner limit $q \rightarrow \infty$ implies $G_0 \rightarrow 0$ leading to a finite conductance maximum of $\Gbg +G_0(1+q^2)$ at $\tilde{\epsilon} = 1/q$.
  
\section{Results}

As mentioned above, Fig.~\ref{figure2}(a) shows three consecutive, well-separated and relatively 
narrow
resonances on a background that varies smoothly in the 
range
0.11 -- 0.22 $e^2/h$.
The 
conductance does not vanish at resonance, as would occur if the destructive
interference between the transmission paths were complete, presumably because of an incoherent component.
The resonances in the center and right are of the Fano form Eq.~(\ref{qformula2}) with 
asymmetry parameters given in the figure caption.

We might imagine that we could calculate the combined transmission through
the resonant and nonresonant channels simply by adding the complex
amplitudes 
\end{multicols}
\widetext
\bef[htb]
\centering
\PSbox{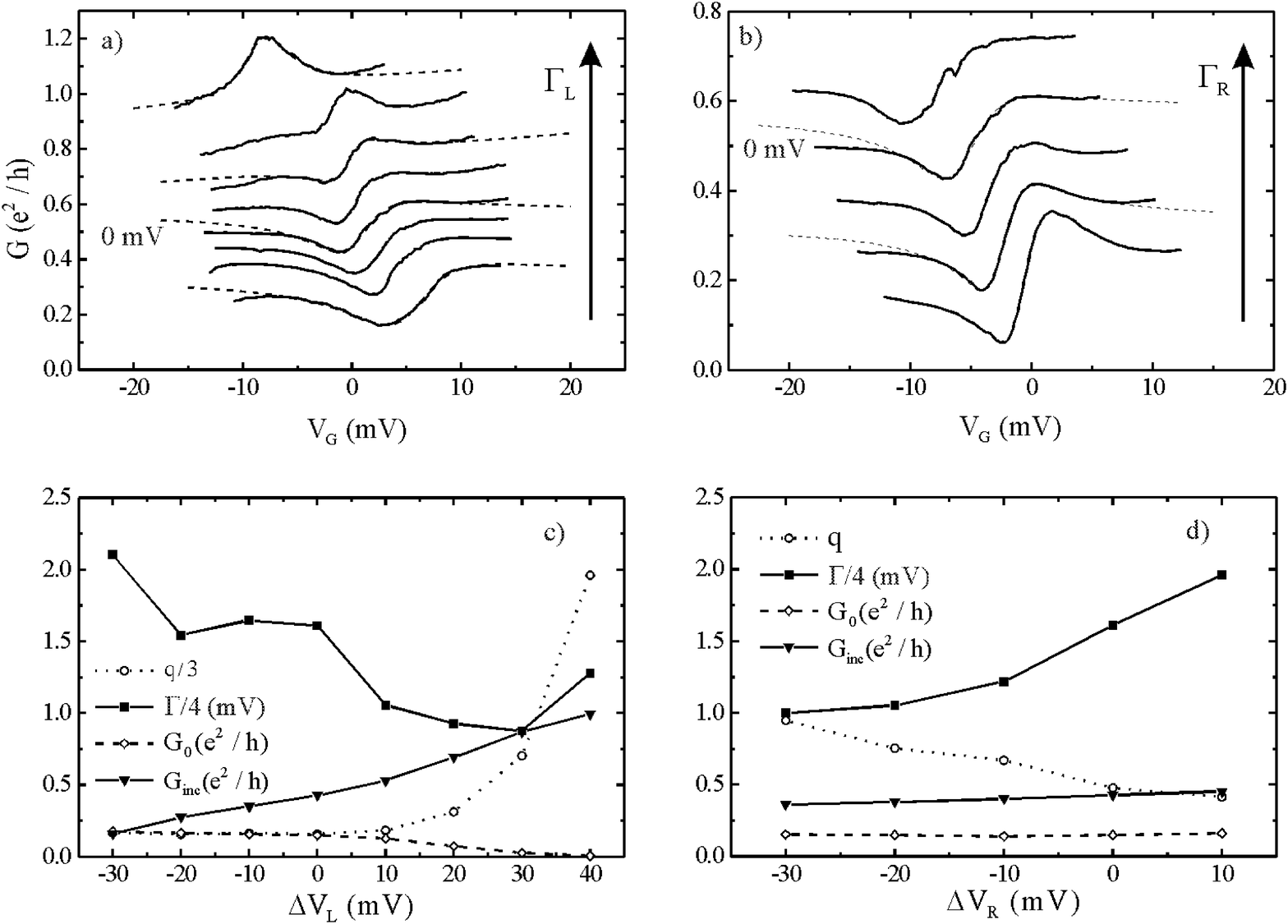 hscale=28 vscale=28}{16.5cm}{11.4cm}
\caption{Conductance resonances as a function of the gate voltage for various
strengths of the
tunnel couplings (a) $\GL$ and (b) $\GR$. The data sets labeled $0\ \mbox{mV}$ are identical for both (a) and (b) and correspond to the case where $-145$ mV is applied to both the left and the right split gate electrodes (labeled I in Fig.~\ref{figure1}). The
couplings are increased in the direction of the arrows by successively
increasing the voltages on the respective electrodes in $10\ \mbox{mV}$ steps. All resonances
are displaced horizontally to account for the capacitative shifts of the
resonance positions caused by the different
applied gate voltages. The resonances
in (b) are displaced vertically for clarity in 0.1 $e^2/h$ steps. This was not
necessary in (a) because of the substantial variation of the background. The dashed curves are obtained by fitting the Fano formula (\ref{qformula2}) to the resonances. The best fitting parameters for the resonances in (a) and (b) are
given in (c) and (d), respectively. The parameter $\Gamma$ is given in $\mbox{mV}$ of gate voltage. Multiplication by $\alpha \approx 0.06$ converts these into energy units (meV).}
\label{figure3} 
\enf
\begin{multicols}{2}
\narrowtext
\noindent for transmission through the two channels, each considered 
individually. In fact, the phase
difference between 
the two transmissions, and hence the degree of asymmetry
of the resonant lineshape, depends on the relative strengths of transmission
through the two channels. 
This means that just by changing the {\em strength} of
resonant transmission we can change the shape of the Fano profile in a way
that would naively seem to require changes in {\em phase} of one or both
transmissions. This effect can be achieved experimentally by varying
the voltages on our point contact electrodes, thereby changing the strength
of transmission through each of the two channels, and generally changing
the ratio of their strengths as well.

The influence of the strength of the tunnel couplings $\GL$ and
$\GR$ on the resonances is shown in Fig.~\ref{figure3}.
All the resonances in this figure can be fit very 
well by the Fano
formula Eq.~(\ref{qformula2}). The fitting parameters for the data in Fig.~\ref{figure3}(a) and (b) are plotted in
Fig.~\ref{figure3}(c) and (d), respectively.
Increasing $\GL$ leads to a more symmetric line shape. At the same time, the incoherent background 
grows strongly, while
the difference between maximum and minimum conductance remains almost constant. 
This is reflected in a strong increase of the asymmetry parameter $q$ accompanied by a decrease in $G_0$ leaving the product $G_0(1+q^2)$ nearly unchanged. Additionally a slight decrease in $\Gamma$ is observed. By contrast, increasing $\GR$ leaves the magnitude of the incoherent transmission constant and decreases the asymmetry parameter. At the same time the resonance is broadened, as reflected by an increase in $\Gamma$. 

Two consecutive Fano resonances with small asymmetry parameters,
resulting in nearly symmetric dips, are shown for a variety of
temperatures in Fig.~\ref{figure4}(a). The increase in 
width of the resonances with increasing temperature is in good agreement with the linear $3.5\  \kB T$ dependence expected from Fermi-Dirac broadening (Fig.~\ref{figure4}(c)). From this we obtain the conversion factor $\alpha = .059 \pm .006$ that relates gate voltage $\Vg$ to energy.  

In contrast to this the temperature dependence of the dip amplitude is not that expected from Fermi-Dirac
\bef[htb]
\centering
\PSbox{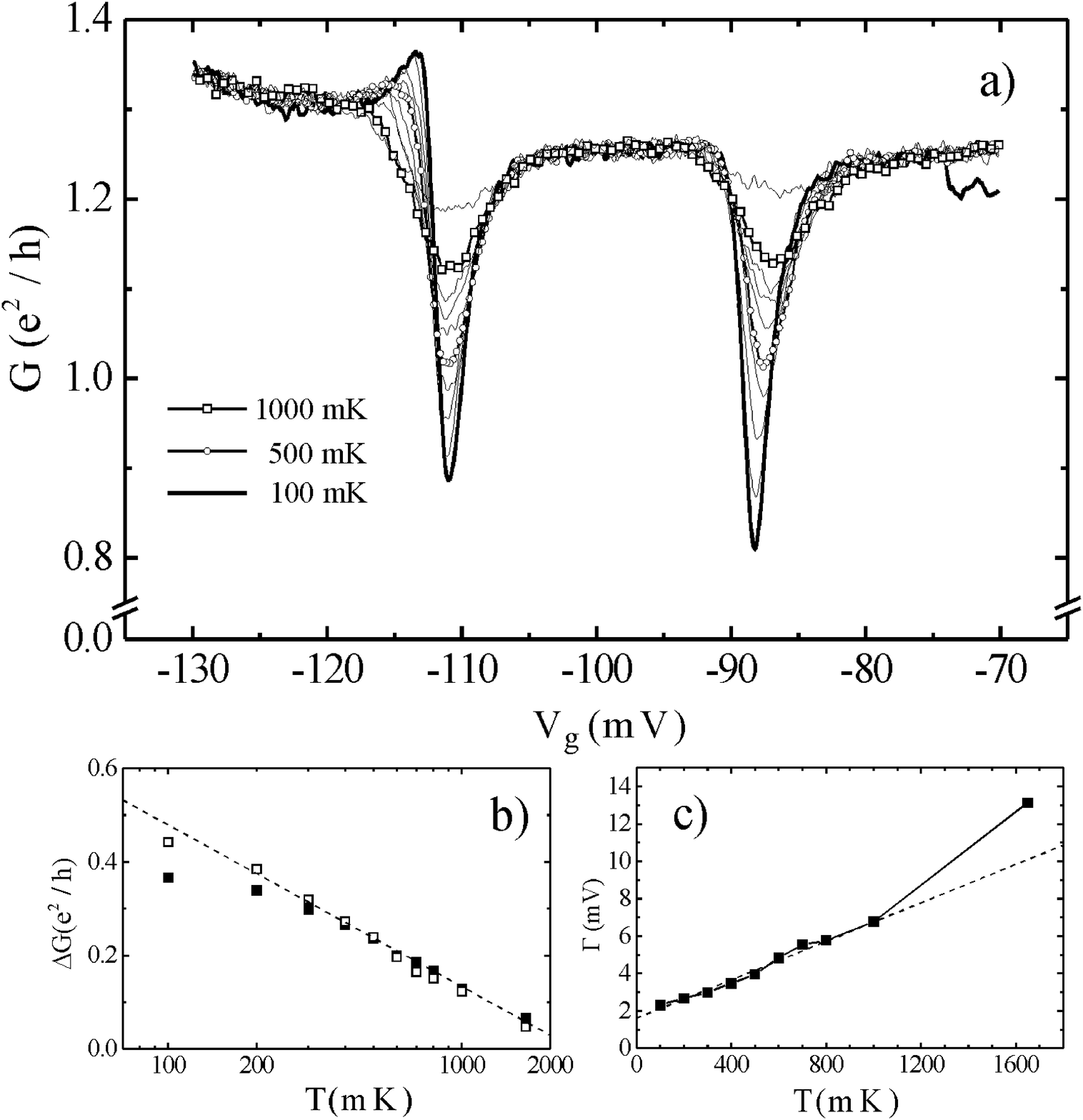 hscale=15 vscale=15}{9.6cm}{8.4cm}
\caption{(a) Temperature dependence of
the conductance for two resonances. At $100\ \mbox{mK}$ the asymmetry parameters for the left and right resonance are $q = -.35$ and $q = -.13$, respectively. The background has been adjusted for a slight increase with temperature, corresponding to only $0.06\ e^2/h$ at the highest temperatures. (b) Dip amplitude with respect to the background conductance as a function of temperature. (c) Width $\Gamma$, measured in $\mbox{mV}$ of gate voltage, of the right resonance as a function of temperature as determined from a fit to the Fano formula Eq.~(\ref{qformula2}). The slope of the linear fit (dashed line) at low temperatures gives a value of $\alpha = .059 \pm .006$ relating gate voltage to energy.}
\label{figure4} 
\enf
\noindent broadening but rather is reminiscent of that seen for peaks in the Kondo regime \cite{davidprl}.
Indeed, the amplitude, measured with
respect to the background, shows a logarithmic dependence on $T$
over almost an order of magnitude as shown in Fig.~\ref{figure4}(b).
In addition, the symmetric dip on the right shows a shift of the
resonance energy with temperature suggesting that the energy is renormalized at
low temperatures just as for conductance peaks in the Kondo
regime. These data resemble curves obtained from a mean-field treatment of quantum interference in the Kondo regime \cite{heemeyer}.

The situation becomes even more intriguing when one examines the
differential conductance as a function of both gate voltage and
source-drain voltage in the vicinity of two dips.
The results of this measurement are shown on a
gray-scale plot in Fig.~\ref{figure5}, where a clear
diamond-shaped structure is traced out by the resonances as one 
varies the two voltages. This behavior is familiar from many experiments
in the Coulomb blockade regime \cite{foxman,kouwenhoven,weis}. Indeed, close
scrutiny reveals additional dips moving parallel to those forming the main
diamonds, analogous to subsidiary peaks seen in the Coulomb blockade regime,
which result from excited states of the electron droplet.  

However, the
results in Fig. \ref{figure5} are different in important ways. The
resonances are dips rather than peaks, and they appear on top of a
continuous background conductance that is almost independent of the 
applied voltages.

From the slopes of the diamonds' boundaries it is possible 
to determine the parameters $\alpha = C_{\mbox{\scriptsize gate}}/C_{\mbox{\scriptsize tot}} = .049 \pm .005$ and $\beta = C_{\mbox{\scriptsize lead}}/C_{\mbox{\scriptsize tot}} = .66 \pm .09$. Here $C_{\mbox{\scriptsize tot}}$ is the total capacitance coupling the electron droplet to its environment whereas $C_{\mbox{\scriptsize gate}}$ is the capacitance only to the plunger gate and $C_{\mbox{\scriptsize lead}}$ the capacitance only to a lead. The value for $\alpha$ is in good agreement with the one obtained from the temperature dependence above. The bottom resonance in Fig.~\ref{figure5} is identical to the left one in Fig.~\ref{figure4} allowing us to determine the  spacing in gate voltage of three successive peaks.  We assume that the smaller spacing is given by $U/\alpha$ and the larger by $(U+\Delta\epsilon)/\alpha$. Using this and the width from Fig.~\ref{figure4}, we find  $U = 1.13 \pm .12\  \mbox{meV}$, $\Gamma = 105 - 120\  \mbox{$\mu$eV}$ and $\Delta\epsilon = .66 \pm  .07\ \mbox{meV}$.
It is surprising that the charging energy is only about $40\%$ smaller than what we find in the Kondo regime.  It is even more surprising that $\Gamma$ has decreased by $50\%$ compared to the Kondo regime, rather than increasing as expected, even though we have opened up the tunnel barriers resulting in a sizable non-resonant
conductance and resonant dips instead of peaks.

Also reminiscent of the Kondo regime are features that remain
pinned near $\Vds = 0$ as the gate voltage is varied, seen as a faint vertical stripe in the center of Figure 5.
In the Kondo regime
this results from a sharp peak in the density of states at the Fermi
energy caused by coupling of the localized spin on the artificial atom 
to the spins in the metallic leads. However, in Fig.~\ref{figure5} the features
do not
\bef[htb]
\centering
\PSbox{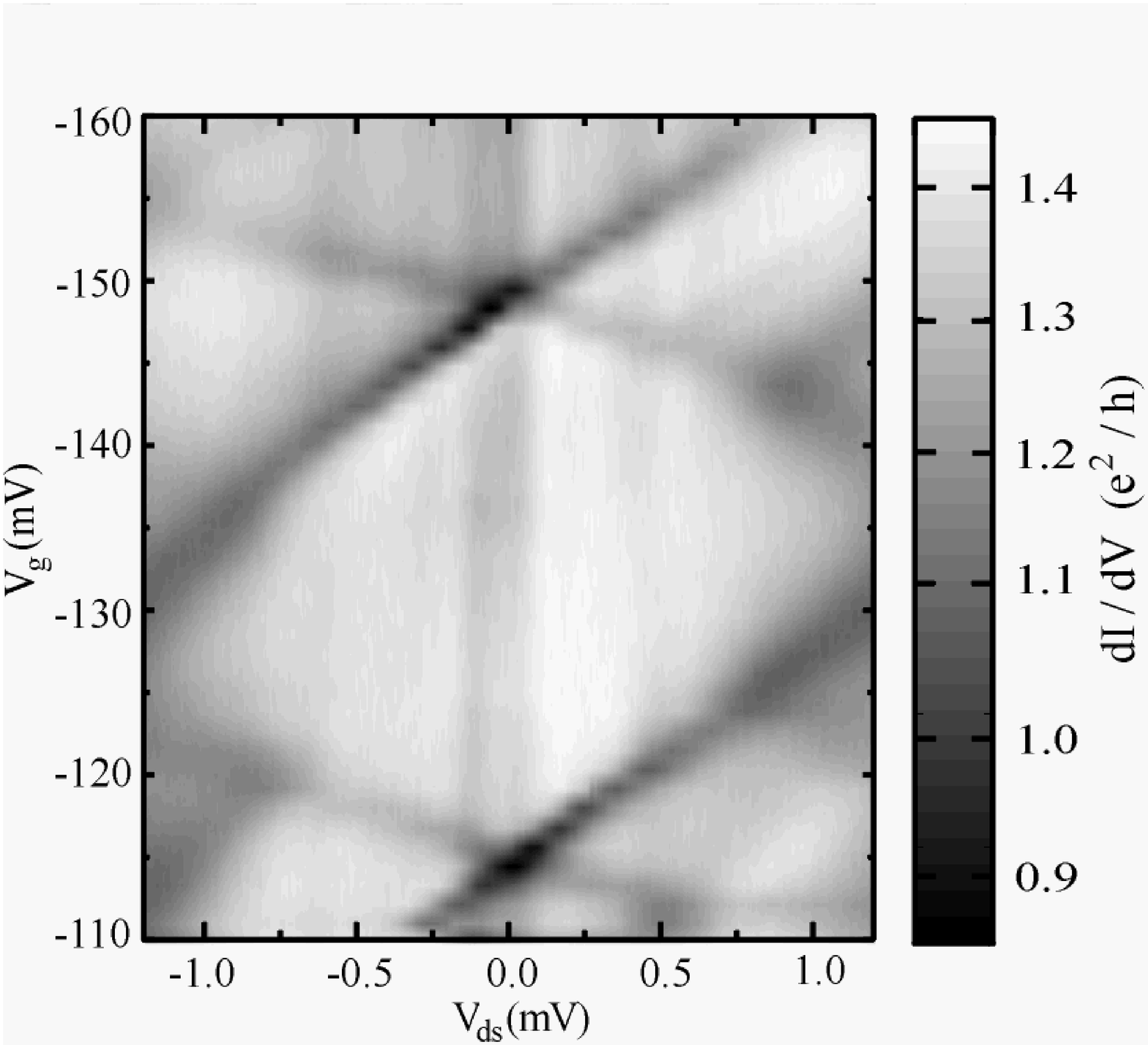 hscale=28 vscale=28}{8cm}{7.7cm}
\caption{Differential conductance ($dI/d\Vds$) at $T = 100\ \mbox{mK}$ as a function of both the
bias voltage $\Vds$ across the SET and the gate voltage
$\Vg$. Notice that the features are produced by dips in the
conductance rather than peaks and that there are weak features near $\Vds = 
0$}
\label{figure5} 
\enf
\bef[htb]
\centering
\PSbox{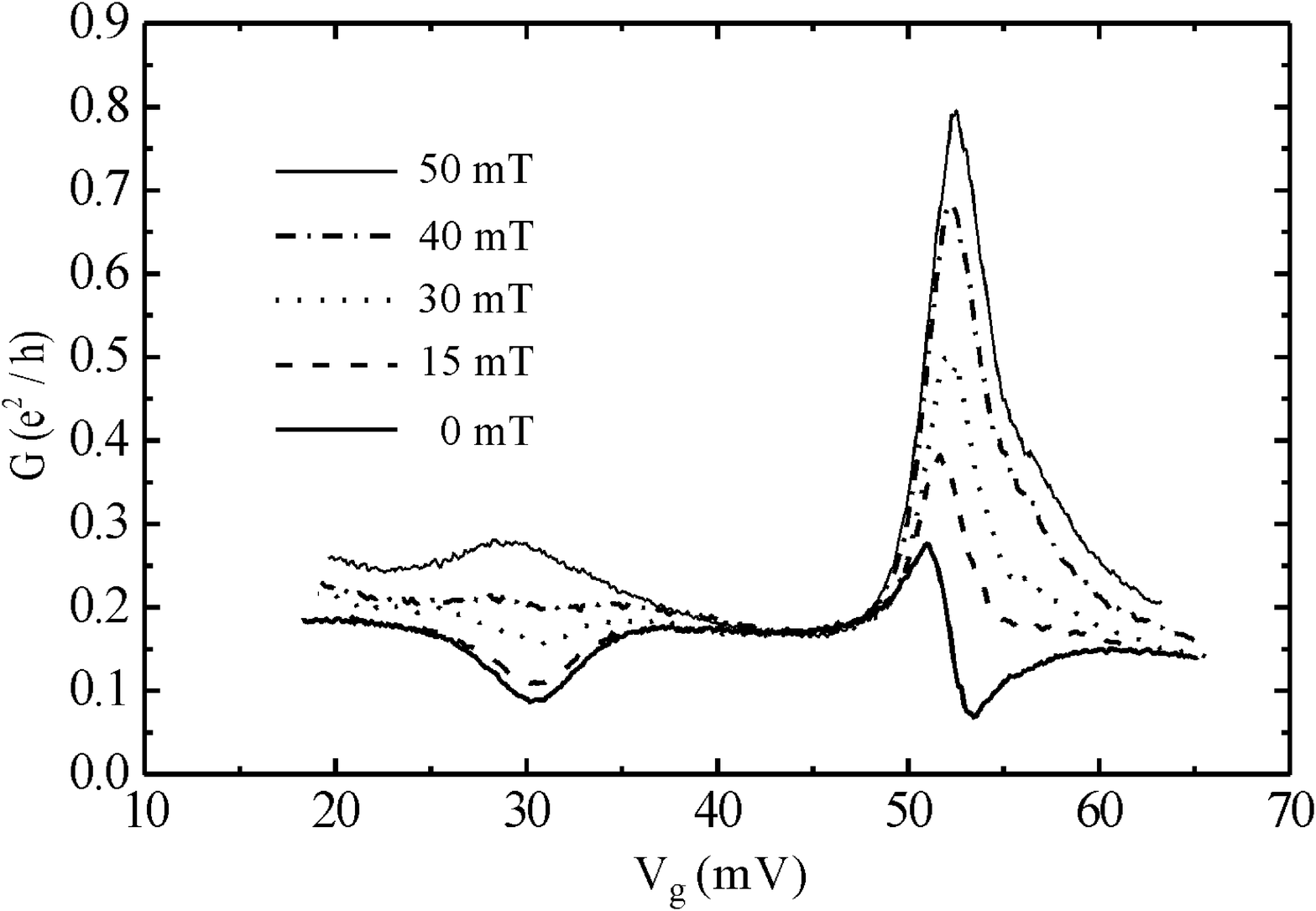 hscale=15 vscale=15}{10cm}{6cm}
\caption{Conductance as a
function of gate voltage for various magnetic fields applied perpendicular
to the 2DEG.}
\label{figure6} 
\enf
\noindent depend on whether an even or odd number of
electrons is on the droplet as evinced by their spanning more than
two resonances. Furthermore, the zero-bias
peak in the measurements of Goldhaber-Gordon $\it{et\ al.}$\cite{davidprl} has
an amplitude that depends strongly on the separation in $\Vg$ from the
resonance, whereas the features in Fig.~\ref{figure5} are nearly
independent of $\Vg$.

The effect of a magnetic field perpendicular to the 2DEG
containing the electron droplet is dramatic (Fig.~\ref{figure6}).
A field of only 15 mT produces a clear effect on the line shape, and a
field of 50 mT completely transforms it from an
asymmetric resonance with a dip into a somewhat asymmetric peak.

\section{Discussion}

\subsection{Nature of interfering channels}

The good fit of the Fano form to our measurements makes it
clear that we are observing interference between resonant and
non-resonant paths through our SET. Were electrons non-interacting,
it would not be surprising for resonant and non-resonant transmission
to coexist. We can see this with the help of a semiclassical non-interacting
analog for the SET: a cavity with two small openings to reservoirs on the left and right sides. Electrons incident on the cavity from the left side at a random angle would bounce around the cavity, only achieving significant transmission
to the right should their energy match an eigenenergy of the cavity. This
process would give resonant transmission. In contrast, electrons incident
on the left side at exactly the correct angle could traverse the cavity and
leave through the righthand opening without suffering any bounces inside the
cavity. This is a nonresonant process, independent of electron energy.
However, in our case the resonant channel shows all the
features typical of charging of artificial atoms: the
near-periodicity in gate voltage of the conductance resonances and the
diamond structure and additional features associated with excited states
in the differential conductance. It is surprising that we see such strong
evidence for charge quantization (Fig.~\ref{figure5}) coexisting with 
a continuous open channel through our SET. We can resolve this seeming
contradiction if two paths for an electron through the SET have
traversal times long and short respectively compared
to $\hbar/U$. The path which spends a long time in the region of the electron
droplet (passing through a well-localized state) must respect the charging
energy, so it exhibits Coulomb blockade resonances in conductance. The other
path can temporarily violate energy conservation, adding a charge to the
electron droplet during a rapid traversal of the SET~\cite{cotunneling}.
We have calculated based on a suggestion by Brouwer {\em et al.} that the
time required for ballistic traversal of our electron droplet should be
comparable to $\hbar/U$~\cite{brouwer}.

A puzzle remains: How can the resonances be quite narrow (Fig.~4(a)) even
though the point contacts are more open than in the Kondo regime (Fig.~2(b)),
as demonstrated by a nonresonant background conductance larger than $e^2/h$.
Conductance through a point contact in the lowest subband cannot exceed
$2e^2/h$, corresponding to a transmission probability of unity. When
electrons can be transmitted across a point contact into two distinct states,
the {\em sum} of the two transmissions can be no greater than unity.
So a large transmission across the first point contact into the delocalized
state which ballistically traverses the electron droplet implies a
correspondingly reduced transmission into the localized, resonant state.
The same analysis holds for electrons traversing the second point contact
to exit the droplet. The width $\Gamma$ of a resonant state is given by
the rate of escape of electrons from that state, which is in turn proportional
to the sum of the transmissions across the two point contacts into or out of
that state. This explains how the presence of a nonresonant transmission
channel can actually make the transmission resonances sharper~\cite{baksheyev}.

We have considered alternative explanations for the origin of the
nonresonant background conductance, and find them unlikely. A path
that circumvents the electron droplet might lead to a continous
contribution to the conductance. However, a parallel conduction path in the
dopant layer above the 2DEG has been ruled out by Hall and
Shubnikov-de Haas measurements on samples from the same wafer with
large-area gate electrodes. Since we observe Fano resonances in each
of several devices we have studied, it also seems unlikely that the effect
is a consequence of channels bypassing the quantum dot in the plane of
the 2DEG, caused by lithographic defects or non-uniform charge
distributions. Most importantly, should a path circumvent the electron
droplet the resulting background would be incoherent. For interference
it is important that both paths include the two point contacts since
only they can act as coherent source and detector.

Detailed measurements of the evolution from the Kondo regime to the
Fano regime are underway, with the hope of further elucidating the nature of
the nonresonant background conductance.

\subsection{Magnetic field dependence}

Changes in transport even at very small field scales are not
unexpected given the droplet's geometry and the 2DEG properties. For
our nonresonant transmission, electrons traverse the droplet directly,
so backscattered paths enclose approximately the area of the
droplet. Assuming an electron droplet of 100 nm diameter, one flux
quantum $\Phi_0 \equiv h/e$ penetrates the droplet at approximately
530 mT applied field. Thus, at this field scale, changes in nonresonant
conductance would result from the breakdown of coherent
backscattering~\cite{serota}. However, the resonant path through
our droplet is less strongly connected to the leads, so electrons
traverse the droplet by more roundabout paths, enclosing more
net area than simply that of the droplet. This means that breakdown of
coherent backscattering should occur at much lower flux through the
droplet, $\Pcorr = \Phi_0/\sqrt{g(\Delta\epsilon/\Gamma)}$, where $g
\equiv G/(2e^2/h)$ is the dimensionless conductance of the droplet
itself~\cite{serota}.  We saw earlier that $\Delta\epsilon/\Gamma
\approx 5$; and $g$ should be comparable to the conductance of the
2DEG $\gd \approx 300$, though somewhat suppressed because the
electron density of the droplet is less than that of the
2DEG~\cite{beenakker}. Taking $g = \gd$, we arrive at 12 mT as
an estimate and lower bound for $\Pcorr$, consistent with our
empirical observations~\cite{aleiner1}.

Empirically (Fig.~\ref{figure6}), small magnetic fields produce
dramatic alterations in the resonances, while leaving the nonresonant
background essentially unchanged.  The argument made above explains
the changes in transport at small magnetic fields as resulting from
the breakdown of coherent backscattering in the resonant channel, and
the concomitant increase in forward transmission through that
channel. Since non-resonant transmission is not affected at these low
fields, the net result is an enhancement of $q$. The alternative
explanation --- that the magnetic field destroys the interference
between non-resonant and resonant processes, transforming resonant
dips into peaks --- does not account for the extremely low field
scale at which the change occurs, nor does it fully match the data.
The breakdown of coherent backscattering indeed is caused by the loss
of the special phase relation between pairs of {\em time-reversed}
paths, but here we are concerned with interference between two
distinct {\em forward scattering} paths (resonant and non-resonant)
which do not form a time-reversed pair.  In addition, interpreting the
zero-field data as the simple interference between two paths would
suggest that the resonant path has half the amplitude of the
non-resonant path. Yet applying a field causes the resonant
contribution of the right-hand peak to exceed the non-resonant
background by a factor of three. Both these considerations lead us to
believe that changes in amplitude (and perhaps phase) for the resonant
process due to applied field are more important than destruction of
coherence by that field.

\section{Acknowledgements}

We acknowledge fruitful discussions with P. Brouwer,
L. I. Glazman, R. J. Gordon,
G. Sch\"{o}n, H. Sch\"{o}ller, S. H. Simon, A. Yacoby and especially J. U.
N\"{o}ckel and Ned Wingreen. J. G. thanks NEC, and D. G.-G. thanks
the Hertz Foundation, for graduate fellowship support. This work
was supported by the US Army Research Office under contract DAAG 55-98-1-0138, and by the
National Science Foundation under grant number DMR-9732579.

\end{multicols}

\begin{thebibliography}{10}
\bibitem{qcomplex}
The incoherent background $\Gbg$ is introduced to account for the non-vanishing conductance at the resonance minimum. We can also account for this behavior by assuming a complex asymmetry parameter $q$ \cite{noeckel,heemeyer}.
\bibitem{cotunneling} This
mechanism is an extension of the well-known cotunneling process
which accounts for Lorentzian tails of Coulomb blockade conductance peaks.
\bibitem{aleiner1}
Refs.~\cite{aleiner} presents
and \cite{marcus} experimentally corroborates a result for $\Pcorr$
slightly different from that derived in ref.~\cite{serota}. Our experiment
cannot distinguish between the two theoretical predictions, which differ
by less than a factor of two for our parameters.
\end{thebibliography}
\end{document}